\documentclass[%
 aip,
 amsmath,amssymb,
 reprint,%
]{revtex4-1}

\usepackage{graphicx}
\usepackage{dcolumn}
\usepackage{bm}

\usepackage[utf8]{inputenc}
\usepackage[T1]{fontenc}
\usepackage{mathptmx}
\usepackage{etoolbox}
\usepackage{braket}
\usepackage{bbold}

\newcommand{\Id}{\mathbb{1}}
\newcommand{\floor}[1]{\lfloor #1 \rfloor}
\newcommand\numberthis{\addtocounter{equation}{1}\tag{\theequation}}

\makeatletter
\def\@email#1#2{%
 \endgroup
 \patchcmd{\titleblock@produce}
  {\frontmatter@RRAPformat}
  {\frontmatter@RRAPformat{\produce@RRAP{*#1\href{mailto:#2}{#2}}}\frontmatter@RRAPformat}
  {}{}
}%
\makeatother
\begin{document}

\preprint{AIP/123-QED}

\title{Efficient ground state preparation in variational quantum eigensolver \\
with symmetry-breaking layers}


\author{Chae-Yeun Park}
\affiliation{Xanadu, M5G 2C8 Toronto, Canada}
\affiliation{Institute for Theoretical Physics, University of Cologne, 50937 K{\"o}ln, Germany}
\email{chae.yeun.park@gmail.com}

\date{\today}

\begin{abstract}
Variational quantum eigensolver (VQE) solves the ground state problem of a given Hamiltonian by finding the parameters of a quantum circuit ansatz that minimizes the Hamiltonian expectation value.
Among possible quantum circuit ans\"{a}tze, the Hamiltonian variational ansatz (HVA) is widely studied for quantum many-body problems as the ansatz with sufficiently large depth is theoretically guaranteed to express the ground state.
However, since the HVA shares the same symmetry with the Hamiltonian, it is not necessarily good at finding the symmetry-broken ground states that prevail in nature.
In this paper, we systematically explore the limitations of the HVA for solving symmetry-broken systems and propose an alternative quantum circuit ansatz with symmetry-breaking layers.
With extensive numerical simulations, we show that the proposed ansatz finds the ground state in depth significantly shorter than the bare HVA when the target Hamiltonian has symmetry-broken ground states.
\end{abstract}

\maketitle

\section{Introduction}

Experimental progress in controlling quantum systems has allowed the first quantum computational advantage claim~\cite{supremacy} in recent years.
This achievement initiated the race between noisy intermediate-scale quantum~\cite{preskill_nisq} (NISQ) computers and the state-of-the-art classical algorithms~\cite{pan2022solving}.
Now, a NISQ device with hundreds of qubits is available~\cite{kim2023evidence}, and those devices are expected to solve a practical computational problem beyond the reach of classical computers.
Among possible applications, variational quantum eigensolver (VQE)~\cite{peruzzo2014variational,mcclean2016theory} that solves the ground state problem of quantum many-body Hamiltonian has gained lots of attention recently (see Ref.~\onlinecite{cerezo2020variational} for a recent review).

The VQE combines a parameterized quantum circuit and a classical optimization algorithm: A quantum circuit evaluates the expectation value of the Hamiltonian and its derivatives for the output quantum state, whereas a classical optimizer finds better parameters that minimize the energy.
As solving the ground state problem of quantum Hamiltonians is difficult for classical computers, one may easily get a possible advantage of VQEs.
Still, it is less understood which ansatz and classical optimizer should be used to solve a given Hamiltonian efficiently.

One of the most widely studied ans\"{a}tze for solving many-body spin Hamiltonians is the Hamiltonian variational ansatz (HVA)~\cite{wecker2015progress,hadfield2019quantum}.
Inspired by a short-depth quantum algorithm for solving combinatory optimization problems~\cite{farhi2014quantum},
the HVA is constructed using rotating gates whose generators are the terms of the Hamiltonian, which resembles the Suzuki-Trotter decomposition~\cite{suzuki1976generalized}.
Even though the HVA solves many different Hamiltonians reliably~\cite{ho2019efficient,wierichs2020avoiding,wiersema2020exploring}, however,
it is not necessarily good at solving problems with a symmetry-broken nature, which prevails in many-body systems.
Indeed, there are local Hamiltonians whose ground states cannot be generated by this ansatz in a constant depth, albeit such a circuit exists, as the circuit obeys the same symmetry as the Hamiltonian~\cite{bravyi2020obstacles}.

In this paper, we devise a symmetry-breaking ansatz and explore its power for solving many-body Hamiltonians.
We construct our ansatz by adding symmetry-breaking layers to the HVA.
The power of those symmetry-breaking ans\"{a}tze is numerically tested using the transverse-field Ising (TFI) and the transverse-field cluster (TFC) models, which have the $\mathbb{Z}_2$ and $\mathbb{Z}_2 \times \mathbb{Z}_2$ symmetries, respectively.
Our results show that the symmetry-breaking ansatz can find the ground state in a constant depth when the Hamiltonian has symmetry-broken ground states, whereas the bare HVA requires a linear depth for solving the same problem.
We also show that, by adding a symmetry penalizing term to the loss function, one can choose a particular symmetry-broken state among degenerate ground states.

\section{Preliminaries}
\subsection{Hamiltonian variational ansatz}
The HVA can be regarded as a parameterized version of the Suzuki-Trotter decomposition.
When the Hamiltonian is written as $H=\sum_{i=1}^\Gamma h^{(i)}$, the HVA is given by
\begin{align}
    \ket{\psi(\theta)} = \prod_{k=1}^{k=d} \Bigl[ \prod_{i=1}^\Gamma \exp(-i h^{(i)} \theta_i) \Bigr] \ket{\psi_0},
\end{align}
where $\ket{\psi_0}$ is the ground state of $H_0$ which is adiabatically connected to $H$.
We note that the HVA is not uniquely defined for a given Hamiltonian, as it depends on how we group the terms of the Hamiltonian.
In this paper, we use the most natural construction that groups commuting terms into one generator, i.e., each $h^{(i)}$ has commuting terms.
The same construction is also used in Refs.~\onlinecite{ho2019efficient,wiersema2020exploring}.

For example, let us consider the Hamiltonian of the transverse field Ising (TFI) model for $N$ qubits, which is given by
\begin{align}
	H_{\rm TFI} = -\sum_{i=1}^N Z_i Z_{i+1} - h \sum_{i=1}^N X_i \label{eq:tfi_ham}.
\end{align}
For $H_0 = -\sum_i X_i$, an adiabatic path between $H_0$ and the TFI given by $H(\tau) = (1-\tau) H_0 + \tau H_{\rm TFI}$ for $0 \leq \tau \leq 1$.
As $\ket{+}^{\otimes N}$ is the ground state of $H_0$ (i.e., at $\tau=0$), the adiabatic evolution is obtained as follows:
\begin{align}
    \ket{\rm GS} &\approx e^{-i \int_{0}^T H(t/T) dt } \ket{+}^{\otimes N }. \label{eq:ground_state_adiabatic}
\end{align}
When $T$ is sufficiently larger than $1/(\Delta E)$, where $\Delta E$ is the energy gap of the system defined as the difference between the ground and the first excited energies of $H(\tau)$, the adiabatic theorem guarantees that the RHS of Eq.~\eqref{eq:ground_state_adiabatic} is close to the true ground state of $H_{\rm TFI}$ (i.e., at $\tau=1$).

We additionally consider the first-order Suzuki-Trotter decomposition of the adiabatic path, which is given by
\begin{align*}
    &e^{-i \int_{0}^T H(t/T) dt } \ket{+}^{\otimes N } \\
    &= \prod_{k=p}^1 \exp\Bigl[ i \bigl\{1-k \Delta \tau (1-h) \bigr\} \Delta \tau \sum_{i=1}^N X_i \Bigr] \\
    &\qquad \times \exp\Bigl[ i k \Delta \tau^2 \sum_{i=1}^N Z_iZ_{i+1} \Bigr] \ket{+}^{\otimes N } + O(\Delta \tau), \numberthis \label{eq:ground_state_trotter}
\end{align*}
where $\Delta \tau = 1/p$ and $p$ is the total trotter steps.
One can see that $p$ controls both the Trotter error and the error from the adiabatic evolution.
Thus, by choosing sufficiently large $p$, one can find the ground state faithfully.

The HVA is constructed by replacing the Trotter time steps in Eq.~\eqref{eq:ground_state_trotter} with parameters, i.e.,
\begin{align}
    &\ket{\psi(\mathbf{\phi}, \mathbf{\vartheta})} \nonumber \\
    &\quad = \prod_{k=D}^1 \exp\Bigl[-i \phi_k \sum_{i=1}^N X_i \Bigr] \exp\Bigl[-i \theta_k \sum_{i=1}^N Z_i Z_{i+1} \Bigr]\ket{+}^{\otimes N }. \label{eq:hva_tfi_primitive}
\end{align}
From the above construction, it is clear that the HVA describes the ground state when $D$ is sufficiently large.

However, it is still unclear whether the depth provided by the HVA is optimal (up to a constant factor) for describing the ground state. 
In this paper, we show that the answer is negative using symmetry-broken systems.

\begin{figure}
    \centering
    \includegraphics[width=0.80\linewidth]{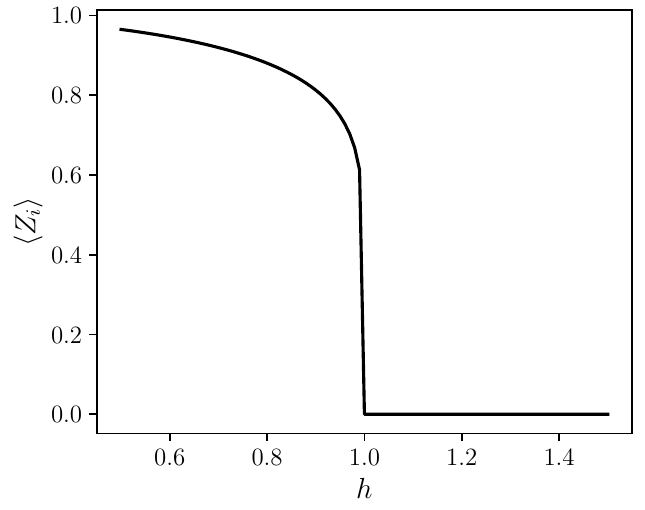}
    \caption{The order parameter $\braket{Z_i}$ as a function of $h$ computed for the ground state of the transverse-field Ising model given by Eq.~\eqref{eq:tfi_ham}. When $h < 1$, there are two ground states, and the order parameter is computed for one of the ground states. Due to the translational invariance, $\braket{Z_i}$ does not depend on $i$. }
    \label{fig:tfi_pt}
\end{figure}

\subsection{Phases related to symmetries}
Quantum many-body systems often break a symmetry characterized by a group.
When it happens, a system has more than two ground states connected by a group generator.
For example, if the $\mathbb{Z}_2$ symmetry is broken, we can find two ground states $\ket{\rm GS_0}$ and $\ket{\rm GS_1}$ such that $\ket{\rm GS_0} = G\ket{\rm GS_1}$ where $G \neq \Id$ is a generator of this symmetry (or a representation of it, more generally) which satisfies $G^2 = \Id$~\cite{wen2004quantum,zeng2019quantum} (see also Ref.~\onlinecite{sbb_lecture_note} for a general introduction).

Conventionally, the symmetry-breaking phenomena are understood using Landau's theory~\cite{landau_symmetry}.
In this theory, symmetry-broken ground states are determined by an order parameter defined by an expectation value of a local observable $O$.
For the $\mathbb{Z}_2$ symmetry-broken ground states, one has an order parameter $O_0 = \braket{{\rm GS_0} | O | {\rm GS_0} }$ and the other has $O_1 = \braket{{\rm GS_1} | O | {\rm GS_1}} = -O_0$.
Contrarily, symmetry-preserving ground states have $\braket{{\rm GS}|O|{\rm GS}} = 0$.
Thus, one can diagnose whether the system breaks the symmetry by computing the order parameter.

A phase transition occurs when the ground state breaks (or restores) a new symmetry as the control parameter of the Hamiltonian changes.
For example, the TFI given in Eq.~\eqref{eq:tfi_ham} has the strength of the external magnetic field $h$ as a parameter.
The Hamiltonian has the $\mathbb{Z}_2$ symmetry under the spin-flip; it does not change under the transformation $Z_i \leftrightarrow -Z_i$.
However, the symmetry of the ground state is broken when $|h| < 1$, i.e., it has two symmetry-broken ground states when $|h| < 1$ in the thermodynamic limit.
In Fig.~\ref{fig:tfi_pt}, we plot the order parameter $\braket{Z_i}$ as a function of $h$~\cite{pfeuty1970one} for one of the ground states.
One finds that $\braket{Z_i}\neq 0$ for $h < 1$, whereas $\braket{Z_i} = 0$ for all $h > 1$.
This implies that a phase transition occurs at $h=1$, and the symmetry is broken when $h < 1$.


In general, true symmetry breaking is only observed in the thermodynamic limit, i.e., a system often has a symmetry-preserving unique ground state for any finite $N$ even when it has degenerated ground states in the thermodynamic limit.
Thus, we need a diagnostic that can tell whether a system breaks symmetry using the ground states for finite-size systems.
Correlation functions are widely used for that purpose.
For example, in 1D system, consider local observables $\{O_i\}$ where each $O_i$ supports site $i$.
Then, the correlation function of the ground state is given by $C(i,j)=\braket{O_i O_j} - \braket{O_i}\braket{O_j}$ where $\braket{\cdot}$ is the expectation value for the ground state from a system of size $N$.
If $C(i,j)$ converges to a constant as $|j-i| \rightarrow \infty$ (we also need to increase $N$ accordingly), the system has symmetry broken ground states in the thermodynamic limit~\cite{sbb_lecture_note}.

With the advent of topological physics~\cite{haldane1983nonlinear,kt_transition,haldane2017nobel}, however, systems with degenerate ground states related by a symmetry that an order parameter or correlation functions cannot diagnose are discovered.
Such an example includes the AKLT model~\cite{affleck1987rigorous}, which has been extensively studied over decades.
The model has two ground states in the open boundary condition, but the correlation function decays exponentially, and none of the local observable can distinguish the two states.
This type of phase is named symmetry-protected topological (SPT) phase~\cite{senthil2015symmetry}.

To incorporate both the conventional (symmetry-broken) and the SPT phases under a single theory, a modern definition of quantum phases has been introduced~\cite{chen2011classification,schuch2011classifying}.
This theory is technically involved, but in essence, it tells that two Hamiltonians ($H_0$ and $H_1$) are in the same phase if there exists a local and gapped parameterized Hamiltonian $H(\tau)$ ($0 \leq \tau \leq 1$) such that  $H(0)=H_0$, $H(1)=H_1$ and $H(\tau)$ has the same symmetry for all $0 \leq \tau \leq 1$ (i.e., there exists an operator $P$ such that $[P,H(\tau)]=0$).
Here, we say $H(\tau)$ is gapped when the difference between the lowest and the first-excited eigenvalues of $H(\tau)$ is bounded below by a constant regardless of the system size $N$.
This definition generalizes Landau's theory since the gap closes in the thermodynamic limit during symmetry-breaking phase transitions.

Since adiabatic evolution of a gapped Hamiltonian is described by a constant-depth quantum circuit~\cite{osborne2007simulating},
the above definition is directly translated into a language of quantum circuits.
Let us assume that Hamiltonians $H_0$ and $H_1$ have ground states $\ket{\psi_0}$ and $\ket{\psi_1}$, respectively.
Then, $\ket{\psi_0}$ and $\ket{\psi_1}$ are in the same SPT phase if there exists a unitary operator $U$ such that (1) $U$ can be expressed by a local and constant depth quantum circuit, (2) $\ket{\psi_1} = U \ket{\psi_0}$, and (3) $[U,P]=0$ where $P$ is the symmetry generator.

If we drop the symmetry restriction from the definition of the SPT phases, we arrive at the definition of topological phases~\cite{bravyi2010topological}.
Namely, two ground states are in different topological phases if we cannot find any local constant-depth circuit $U$ that connects the two states.
It is also known that true topological phases do not exist in the one-dimensional spin systems, i.e., one can always find a constant-depth circuit connecting between ground states of any gapped 1D local Hamiltonians~\cite{verstraete2005renormalization,chen2011classification,schuch2011classifying,bachmann2012automorphic}.
In the following section, we will use such a distinction between the SPT and topological phases to design an efficient quantum circuit ansatz.

\section{Symmetry breaking ansatz}\label{sec:ansatz}
The first model we study is the TFI model for $N$ qubits defined in Eq.~\eqref{eq:tfi_ham}, where we impose the periodic boundary condition $Z_{N+1} = Z_1$.
This model has two distinct (ferromagnetic and paramagnetic) phases depending on the strength of $h$ protected by the spin-flip symmetry $P =\prod_i X_i$~\cite{chen2011complete}. 
The critical point of this model $h=1$ is well known~\cite{pfeuty1970one}.
This implies that if there is a circuit that commutes with $P$ and connects two ground states in different phases (i.e., ground states for $h>1$ and $h<1$), the circuit depth must be larger than a constant~\cite{chen2011classification}.
On the other hand, a finite-depth circuit that connects two different ground states exists if we do not restrict such symmetry to a circuit since the system is gapped unless $h=1$~\cite{chen2011classification,bachmann2012automorphic}.

As the HVA only utilizes terms within the Hamiltonian, it fails to represent such a circuit.
To understand this limitation, let us rewrite the HVA for the TFI given in Eq.~\eqref{eq:hva_tfi_primitive} as follows:
\begin{align}
    \ket{\psi(\mathbf{\phi}, \mathbf{\vartheta})}&=\prod_{k=D}^1 \Bigl[ \mathcal{L}_{x}(\phi_k) \mathcal{L}_{zz}(\vartheta_{k}) \Bigr] \ket{+}^{\otimes N} \label{eq:hva_tfi}
\end{align}
where each layer is given by
\begin{align}
    \mathcal{L}_{x}(\phi) &= \exp[-i\phi\sum_{i=1}^N X_i] \\
    \mathcal{L}_{zz}(\vartheta)&=\exp[-i\vartheta\sum_{i=1}^{N} Z_{i}Z_{i+1}].
\end{align}
Let us denote $L$ by the total number of layers, i.e., $L=2D$ for Eq.~\eqref{eq:hva_tfi}. 
As all gates commute with $P$, i.e. $[P, \mathcal{L}_{zz}] = [P, \mathcal{L}_{x}] = 0$, and the input state $\ket{+}^{\otimes N}$ is the ground state of the Hamiltonian when $h \rightarrow \infty$, we know that preparing the ground state for $h < 1$ requires circuit depth larger than a constant.
Indeed, theoretical and numerical studies have found that this type of ansatz needs depth $D \geq N/2$ to prepare the ground state faithfully~\cite{mbeng2019quantum,ho2019efficient}.

We now add symmetry-breaking layers in VQE ansatz and see whether it can achieve lower circuit depth for preparing the ground state.
Our ansatz for the TFI is given as
\begin{align}
	\ket{\psi(\pmb{\theta})} = \prod_{j=D}^1 \mathcal{L}_z(\phi_j) \mathcal{L}_x(\kappa_j) \mathcal{L}_{zz}(\vartheta_j)  \ket{+}^{\otimes N} \label{eq:hva_tfi_symm_breaking}
\end{align}
where $\mathcal{L}_{z}(\phi_j) = \exp[-i\phi_{j}\sum_{i} Z_{i}]$, and $\pmb{\theta} = \{\vartheta_j, \kappa_j, \phi_j\}_{j=1}^D$ is a set of all parameters.
All layers in the ansatz preserve the translational symmetry, but the $\mathcal{L}_z(\phi_j)$ layers break the symmetry $P$ of the Hamiltonian.
In addition, as each block now has $3$ layers, the total number of layers is $L=3D$.

Symmetry-preserving/breaking ans\"{a}tze with the same $D$ [Eqs.~\eqref{eq:hva_tfi} and \eqref{eq:hva_tfi_symm_breaking}] have different numbers of total gates.
However, the number of entangling gates of both circuits is the same.
When implemented on NISQ devices, the latter dominates the quality of the output states since the expected fidelity of entangling gates is much lower than single-qubit gates (see, e.g., Refs.~\onlinecite{harty2014high,gaebler2016high}).
Thus, in the following discussion, we mainly compare the results from these ans\"{a}tze with the same $D$.

To observe the effect of symmetry-breaking layers, we simulated the VQE for the TFI using the bare HVA and our symmetry-breaking ansatz given in Eq.~\eqref{eq:hva_tfi_symm_breaking}.
We optimize parameters of the ansatz using the quantum natural gradient~\cite{mcardle2019variational,stokes2020quantum,hackl2020geometry}: For each epoch $t$, we update parameters as $\pmb{\theta}_{t+1} = \pmb{\theta}_t - \eta (\mathcal{F} + \lambda_t \Id)^{-1} \nabla_\theta \braket{\psi(\pmb{\theta}_t)|H|\psi(\pmb{\theta}_t)}$ where $\eta$ is the learning rate, $\mathcal{F} = (\mathcal{F}_{ij})$ is the quantum Fisher matrix, and $\lambda_t$ is a (step dependent) regularization constant. 
We choose this optimizer as it works more reliably in solving the ground problem both for classical neural networks~\cite{park2020geometry} and VQEs~\cite{wierichs2020avoiding}.
For the quantum Fisher matrix, we use the centered one $\mathcal{F} = \mathcal{F}_{ij}^{\rm c} = \mathfrak{ Re}\{\braket{\partial_{\theta_i} \psi(\pmb{\theta}_t) | \partial_{\theta_j} \psi(\pmb{\theta}_t)} - \braket{\partial_{\theta_i} \psi(\pmb{\theta}_t) | \psi(\pmb{\theta}_t)}\braket{\psi(\pmb{\theta}_t) | \partial_{\theta_j} \psi(\pmb{\theta}_t)}\}$~\cite{stokes2020quantum} mostly (unless otherwise stated), but the uncentered one $\mathcal{F}_{ij}^{\rm nc} = \mathfrak{ Re}\{\braket{\partial_{\theta_i} \psi(\pmb{\theta}_t) | \partial_{\theta_j} \psi(\pmb{\theta}_t)} \}$~\cite{mcardle2019variational} is also considered when it improves the performance.
The difference between the two can be understood using the notion of the projected Hilbert space~\cite{hackl2020geometry}.
For the hyperparameters, we typically use $\eta = 0.01$ and $\lambda_t = \max(100.0 \times 0.9^t, 10^{-3})$.

\begin{figure}[t]
	\centering
	\includegraphics[width=0.85\linewidth]{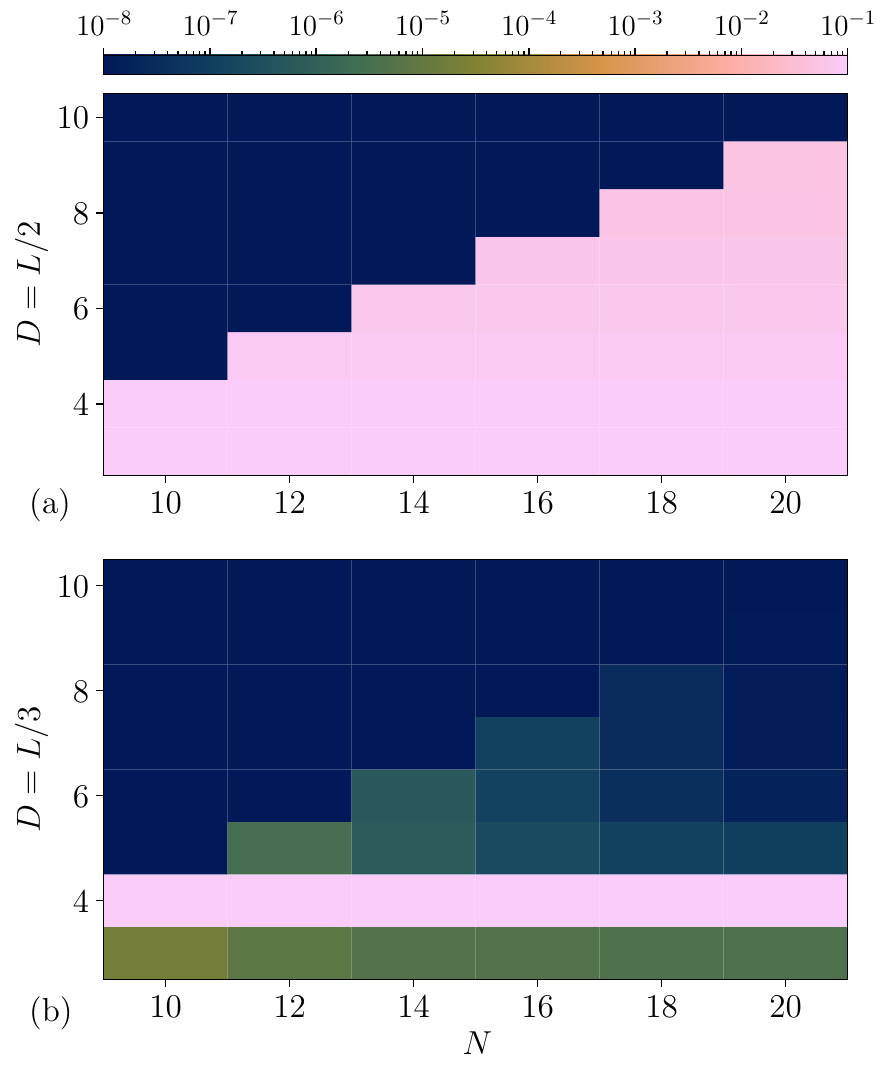}
	\caption{
		\label{fig:ti_tfi}
		Converged normalized energy $\widetilde{E}=(E_{\rm VQE}-E_{\rm GS})/E_{\rm GS}$ as a function of system size $N$ and the number of blocks $D$ (a) without and (b) with symmetry breaking layers for the transverse field Ising (TFI) model where $E_{GS}$ is the true ground state energy.
		We have run $12$ independent VQE runs and taken the best-optimized energy for each $N$ and $D$.
		Initial parameters are sampled randomly from the normal distribution $\mathcal{N}(0, \sigma^2)$ besides $D=3,5$ at (b) where we have added $2\pi/D$ to the parameters of the symmetry breaking layers (see main text for details).
        Note that the total number of layers $L$ is $2D$ for the bare HVA but $3D$ for our symmetry-breaking ansatz.
	}
\end{figure}

\begin{figure}[t]
	\centering
	\includegraphics[width=0.85\linewidth]{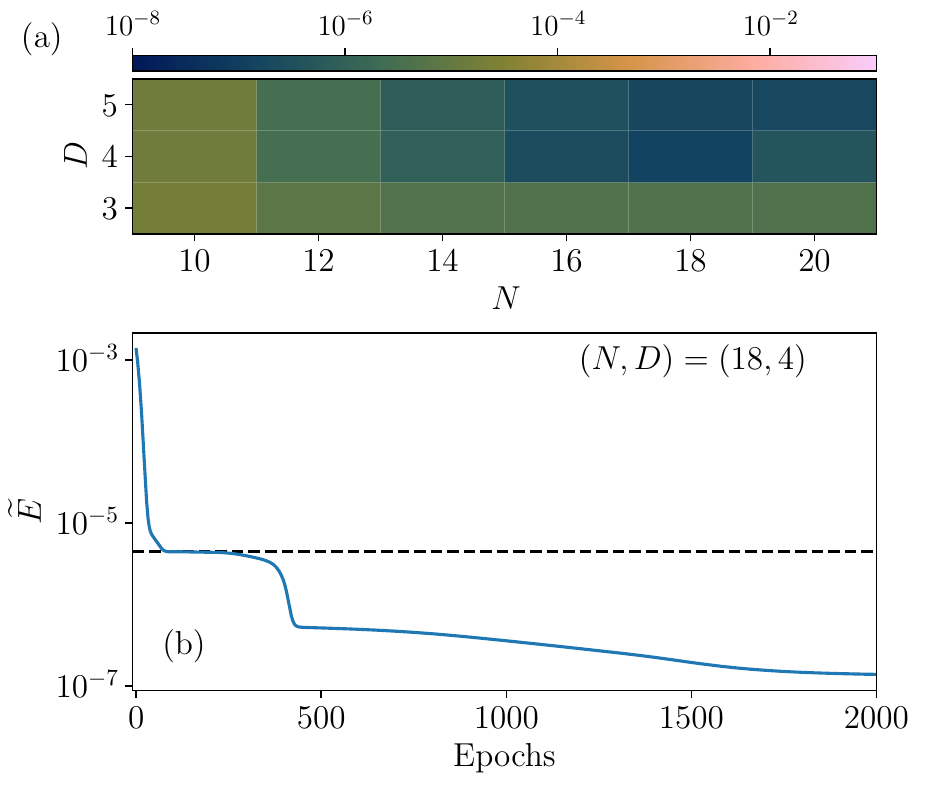}
	\caption{
		\label{fig:tfi_transfer}
		(a) Converged normalized energies $\widetilde{E}$ for system size $N$ and circuit depth $D$ (a) from the transfer learning. Results for $D=3$ is from the random initialization (which are shown in Fig.~\ref{fig:ti_tfi}). For $D=4$, we have trained a circuit after adding a randomly initialized block in the middle of the converged circuit from $D=3$ (for the same $N$). Likewise, we obtained results for $D=5$ using the converged circuits from $D=4$. (b) A learning curve from the transfer learning when $(N,D)=(18,4)$. The dashed line indicates the converged energy from $(N,D)=(18,3)$. 
	}
\end{figure}

We present the optimized normalized energies $\widetilde{E}=(E_{\rm VQE}-E_{\rm GS})/E_{\rm GS}$ for different $N$ and $D$ in Fig.~\ref{fig:ti_tfi} when $h=0.5$. 
We have used the Fisher matrix with the centering term $\mathcal{F}^{\rm c}$, and initial values of parameters $\{\vartheta_j, \kappa_j\, \phi_j\}$ are sampled from the normal distribution $\mathcal{N}(0,\sigma^2)$ with $\sigma = 0.001$ besides $D=3,5$ at Fig.~\ref{fig:ti_tfi}(b).
When $D=3$ (all $N$) and $5$ (for $N \geq 12$), this initialization does not give any better energy than the ansatz without symmetry-breaking layers.
We instead found that initializing parameters for the symmetry breaking layers $\{\phi_j\}$ with samples from $\mathcal{N}(2\pi/D, \sigma^2)$ finds better optima in these cases.
However, this initialization does not change the results for $D=4$ and performs even worse when $D\geq 6$ (see Appendix~\ref{app:optimization_sym_brk} for detailed comparisons).

Fig.~\ref{fig:ti_tfi}(a) shows that the ground state is only found for $D \geq N/2$ when symmetry-breaking layers are absent, which is consistent with Refs.~\onlinecite{ho2019efficient,wierichs2020avoiding}.
On the other hand, results with symmetry-breaking layers [Fig.~\ref{fig:ti_tfi}(b)] clearly demonstrate that converged energies are significantly improved for $5 \leq D < N/2$.
Most importantly, converged normalized energies are $\leq 10^{-7}$ for all $N$ when $D \geq 9$, which implies that our ansatz finds a constant-depth circuit for solving the ground state.
In addition, the results show that there is a finite-size effect up to $N=18$ where the accurate ground state is only obtained when $D \geq N/2$.

However, the converged energies and the true ground state show a large discrepancy when $D=4$.
In fact, the results for $D=4$ are even worse than those of $D=3$, which signals that the optimizer gets stuck in local minima. This type of convergence problem is already observed in Ref.~\onlinecite{wierichs2020avoiding}.
To obtain a better convergence for $D = 4$, we employ a transfer learning technique.
Instead of starting from a randomly initialized circuit, we insert a block in the middle of the converged circuit from $D=3$ and perturb all parameters by adding small numbers sampled from $\mathcal{N}(0, \sigma'^2)$ (where we typically use $\sigma'=0.01$). We then optimize the full circuit using the quantum natural gradient.

\begin{figure}[t]
	\centering
	\includegraphics[width=0.85\linewidth]{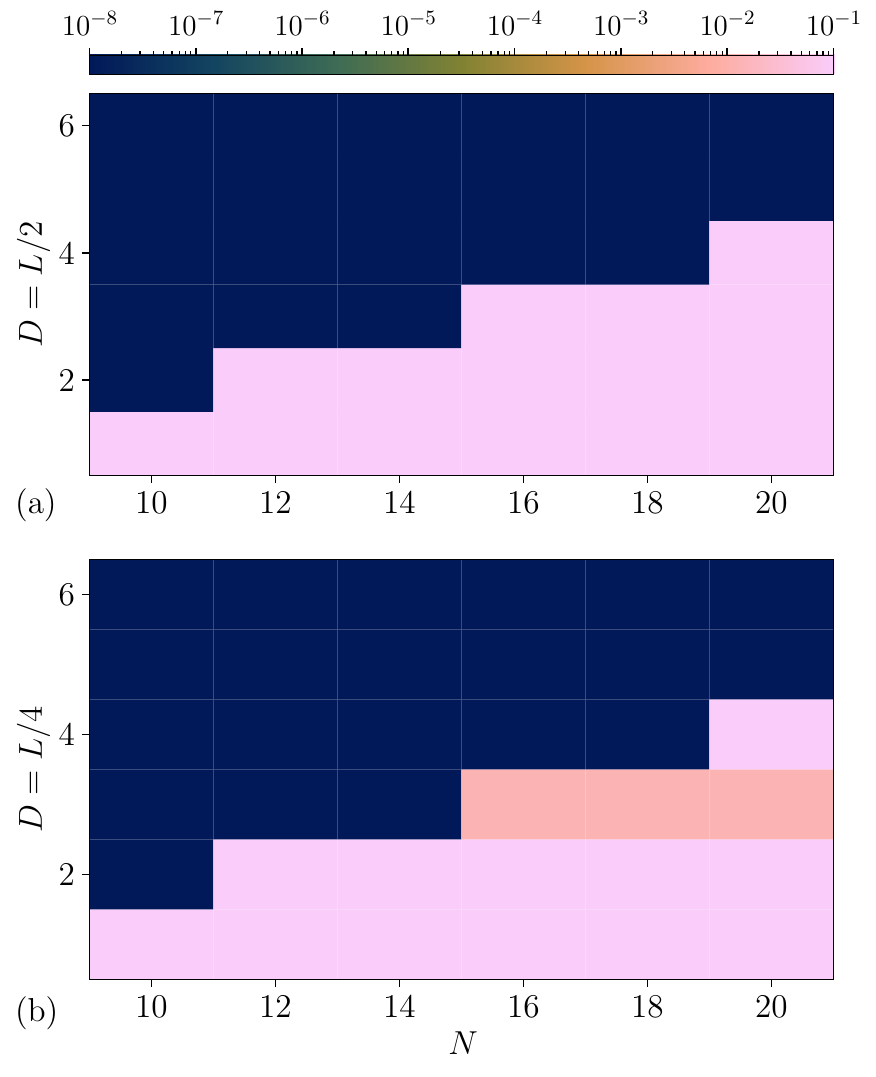}
	\caption{
		\label{fig:ti_cluster}
		Converged normalized energies $\widetilde{E}$ for system size $N$ and the number of blocks $D$ (a) without and (b) with symmetry breaking layers for the transverse field Cluster (TFC) model.
	}
\end{figure}

\begin{figure}[t]
	\centering
	\includegraphics[width=0.85\linewidth]{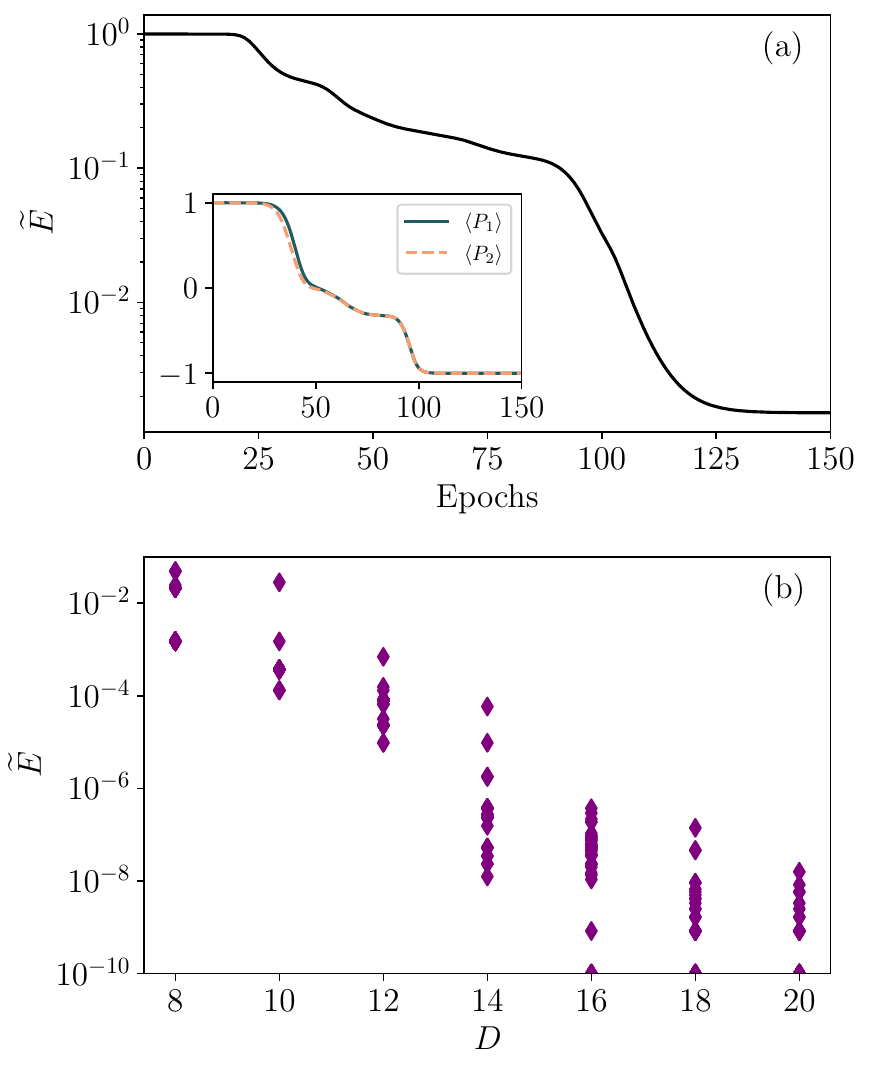}
	\caption{
		\label{fig:cluster_stab}
		(a) Learning curve of the VQE for preparing the ground state with $\braket{P_1} = \braket{P_2} = -1.0$. A circuit with $(N,D)=(14,8)$ is used. Inset shows the learning curves for $\braket{P_1}$ and $\braket{P_2}$. (b) Converged normalized energies $\widetilde{E}$ for difference circuit depths $D$. For each $D$, we show converged energies from $48$ independent VQE runs.
	}
\end{figure}

We show converged energies from the transfer learning in Fig.~\ref{fig:tfi_transfer}. 
Fig.~\ref{fig:tfi_transfer}(a) illustrates that the transfer learning succeeds in finding a better optimum for $D=4$. However, the result for $(N,D)=(10,5)$ is worse than that of the random initialization, which implies that the transfer learning does not necessarily find the global optima.
We additionally plot a learning curve for $(N,D)=(18,4)$ in Fig.~\ref{fig:tfi_transfer}(b) where the initial energy is much higher due to an added perturbation but it eventually finds a quantum state with lower energy.

We next consider the transverse-field cluster (TFC) model, the Hamiltonian of which is given as 
\begin{align}
	H_{\rm TFC} = - \sum_{i=1}^N Z_i X_{i+1} Z_{i+2} - h \sum_{i=1}^N X_i \label{eq:tfc_ham}.
\end{align}
It is known that the phase transition of this model also takes place at $h=1$~\cite{zonzo2018n}.
As the terms $Z_i X_{i+1} Z_{i+2}$ are mutually commuting, the ground state at $h=0$ is the common eigenvector of those operators, which is also known as the cluster state.
Two relevant symmetries that determine the ground state of this model are $P_1 = \prod_{i=1}^{N/2} X_{2i}$ and $P_2 = \prod_{i=1}^{N/2} X_{2i-1}$~\cite{son2011topological,zeng2019quantum}.
By the definition of SPT, a constant-depth circuit that brings the product state $\ket{+}^{\otimes N}$ to the ground state of the Hamiltonian for $h < 1$ does not exist as long as the circuit commutes with $P_1$ and $P_2$.

We study whether a symmetry-breaking layer can improve it. To see this, we construct our ansatz
\begin{align}
	\ket{\psi(\pmb{\theta})} &= \prod_{j=D}^1 \mathcal{L}_z^{\rm even}(\phi_j) \mathcal{L}_z^{\rm odd}(\chi_j) \mathcal{L}_x(\kappa_j) \mathcal{L}_{zxz}(\theta_j) \ket{+}^{\otimes N} \label{eq:cluster_symbrk_ansatz}
\end{align}
where $\mathcal{L}_{zxz}(\theta_j) = \exp[-i \theta_j \sum_{i=1}^N Z_i X_{i+1} Z_{i+2}]$, $\mathcal{L}_z^{\rm even/odd}(\phi_j) = \exp[-i \phi_j \sum_{i=1}^{N/2} Z_{2i/2i+1}]$.
Converged normalized energies with and without symmetry-breaking layers for $h=0.5$ are shown in Fig.~\ref{fig:ti_cluster}.
The results without symmetry-breaking layers (without $\mathcal{L}_{z}^{\rm even/odd}$) show that circuits with $D\geq \floor{N/4}$ solve the ground state accurately.
However, the symmetry-breaking layers only improve the results for $D=3$ when $N \geq 16$ for this Hamiltonian. 
In addition, such an improvement is observed only when the uncentered Fisher matrix $\mathcal{F}^{\rm nc}$ and the initial value $2 \pi /D$ for symmetry-breaking layers are used.

In summary, our ansatz does not find a constant-depth circuit for solving the ground state of the TFC based on the results up to $N = 20$.
We attribute this to the finite-size effect and expect that a much larger system size $N$ (beyond our computational capacity) is required to take advantage of symmetry-breaking ansatz.

\section{Preparing a symmetry broken ground state}
We next consider the cluster Hamiltonian [Eq.~\eqref{eq:tfc_ham} with $h=0$] with the open boundary condition, which is given as
\begin{align}
	H_{\rm cluster} = - \sum_{i=1}^{N-2} Z_i X_{i+1} Z_{i+2}.
\end{align}
Ground states of this model are stabilized by $N-2$ terms ($Z_i X_{i+1} Z_{i+2} \ket{\psi_{\rm GS}}=\ket{\psi_{\rm GS}}$ for $i \in [1,\cdots,N-2]$), so $4$-fold degenerate.
As the operators $P_1$ and $P_2$ commute with all stabilizers, they further define the ground state manifold. 
For the HVA given in Eq.~\eqref{eq:cluster_symbrk_ansatz}, the output state must be $+1$ eigenstate of $P_1$ and $P_2$ as they commute with the circuit and $P_{1,2}\ket{+}^{\otimes N} = +1 \ket{+}^{\otimes N}$, i.e., it can only find the ground state with $P_1=P_2=1$.
In contrast, we show our ansatz [Eq.~\eqref{eq:cluster_symbrk_ansatz}] (with $\mathcal{L}_{zxz}(\theta_j)$ in the open boundary condition) can be used to prepare a particular state within the manifold.

\begin{figure*}[t]
	\centering
	\includegraphics[width=0.85\linewidth]{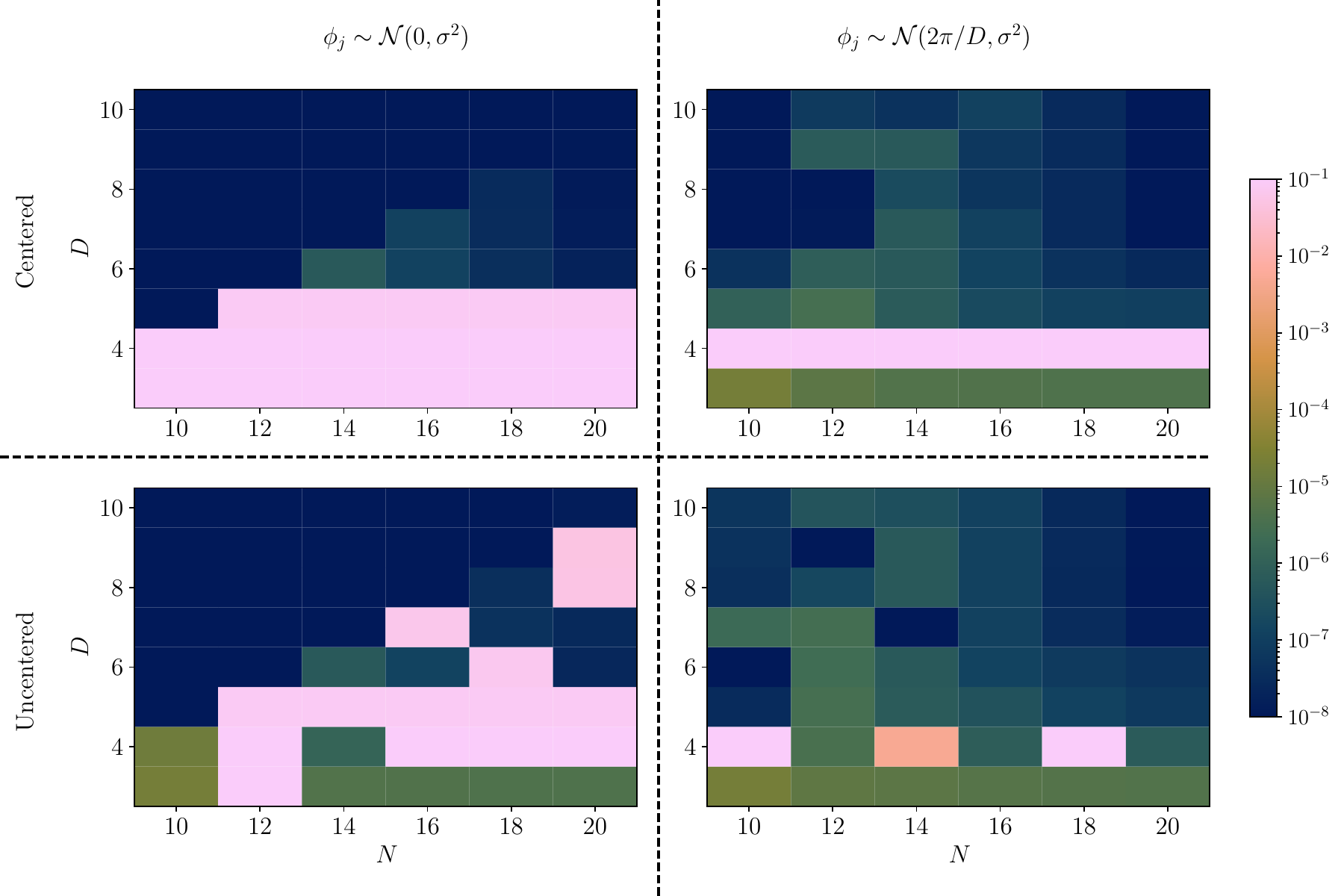}
	\caption{
		\label{fig:tfi_brksym_compare}
		Converged normalized energies $\widetilde{E}$ for the TFI with different optimization set-ups. We have used the centered $\mathcal{F}^{\rm c}$ and uncentered Fisher matrix $\mathcal{F}^{\rm nc}$ for the upper and lower rows, respectively. The left and right columns show results from different initial values for the symmetry-breaking layers. For each $N$ and $D$, we have taken the best-optimized energy from $12$ independent VQE runs.
	}
\end{figure*}

The main idea is minimizing the expectation value of $O=H_{\rm cluster} + \alpha_1 P_1 + \alpha_2 P_2$ for suitable $\alpha_1$ and $\alpha_2$, instead of the Hamiltonian itself. We expect the obtained state to be an eigenstate with the corresponding eigenvalue $\pm 1$ of $P_i$ when $\mathrm{sign}(\alpha_i) = \mp 1$.

For example, we study the VQE for preparing the ground state with $P_1=P_2=-1$ using $(N,D)=(14,8)$.
We have tested VQEs with the varying learning rates $\eta \in [0.01, 0.025, 0.05, 0.1]$ and $\alpha_1=\alpha_2=\alpha \in [0.1, 0.3, 0.5, 1.0, 2.0, 4.0]$, and found that the desired state can be prepared by the best for $\eta=0.025$, $\alpha=2.0$. Especially, all VQE runs have converged to $\braket{P_1}=\braket{P_2}=-1$ (accuracy within $10^{-8}$) with these values of $\eta$ and $\alpha$.
The resulting learning curve is shown in Fig.~\ref{fig:cluster_stab}(a).
However, the converged energy $\widetilde{E} \approx 1.502 \times 10^{-3}$ for $D=8$ is significantly worse than the results from other models we have studied above (we still note that this is comparable to results from the Adam optimizer which gives normalized energies $\approx 10^{-2}-10^{-4}$)~\cite{wiersema2020exploring}. 
We thus further study whether increasing $D$ helps the convergence in Fig.~\ref{fig:cluster_stab}(b).
The results show that the converged energies are getting more accurate as we increase $D$. Precisely, $47$ instances out of $48$ independent VQEs runs have converged to $\widetilde{E} \leq 10^{-8}$ when $D=20$.
This contrasts the barren plateaus narrative~\cite{mcclean2018barren} that more expressive quantum circuits are prone to suffer the optimization problem~\cite{holmes2022connecting}.
We believe that the behavior of our circuits is related to the known fact in classical machine learning that over-parameterized networks converge better~\cite{du2018gradient}, but we leave a detailed study for future work.

\section{Conclusion and outlook}
We studied the limitation of the bare HVA for solving a system with symmetry-broken ground states and proposed an ansatz with symmetry-breaking layers to overcome this problem. 
Such a symmetry-breaking ansatz could enhance the overall performance of the VQE.
Especially for the transverse-field Ising model, we observed that our symmetry-breaking ansatz finds a constant-depth circuit for the ground state, whereas the bare HVA requires the depth to be linear in the system size.
We further proposed a technique to choose a specific symmetry broken ground state among possible ground states.
This was possible by adding a penalizing term to the cost function.

We expect that one can observe similar behavior for continuous symmetry, e.g., $U(1)$ or $\mathrm{SU}(2)$.
However, given that those symmetries are not broken in 1D~\cite{wagner2010mermin}, we should use 2D models to observe the effect of symmetry-breaking layers. 
Symmetry-broken 2D spin models are also physically interesting as they are related to the lattice gauge theory~\cite{byrnes2006simulating}.
However, we leave the detailed study of 2D models for future work, as numerical simulation of such models is more demanding.

Our symmetry-breaking ansatz can be implemented on real quantum hardware.
In this case, one should decompose a gate into the set of gates supported by a target quantum device.
For example, a gate $\mathcal{L}_{zxz}(\theta_j)$, used for the transverse-field cluster model is not naturally supported by most quantum computing architectures~\cite{haffner2008quantum,saffman2010quantum,devoret2013superconducting,henriet2020quantum,tzitrin2020progress}~\footnote{In contrast, all gates for the transverse-field Ising model are naturally supported by most architectures}.
Instead, it can be decomposed into, e.g., $e^{-i\theta Z_{i-1} X_{i} Z_{i+1}} = \mathsf{CZ}_{1,2}\mathsf{CZ}_{2,3}e^{-i\theta X_{i}}\mathsf{CZ}_{1,2}\mathsf{CZ}_{2,3}$, which may increase overall circuit depths.
Still, as our symmetry-breaking ansatz can provide a \textit{scaling} advantage, there is a certain number of qubits that our circuit can solve the problem in a shorter depth than the bare HVA even after such a decomposition.

Incoherent noise can be a significant problem when implemented on noisy quantum hardware.
While a weak incoherent noise may help to find a symmetry-broken ground state when the target state is not entangled~\cite{yamamoto2017coherent}, we generally expect that the noise destroys the quantumness of the circuit, and the circuit outputs a decohered state that is far from a true ground state~\cite{stilck2021limitations}.
As the circuit depth is the main factor limited by this type of noise, our symmetry-breaking ansatz may again have an advantage over the bare HVA when the noise is considered.

\begin{acknowledgments}
The author thanks David Wierichs for helpful discussions.  
This project was funded by the Deutsche Forschungsgemeinschaft 
under Germany's Excellence Strategy - Cluster of Excellence  Matter and  Light for Quantum Computing (ML4Q) EXC 2004/1-390534769 and within the CRC network TR 183 (project grant 277101999) as part of project B01.
The numerical simulations were performed on the JUWELS cluster at the Forschungszentrum Juelich. 
\end{acknowledgments}

\section*{Data Availability Statement}
The source code for the current paper can be found in the Github repository~\cite{efficient-vqe-repo}.

\appendix

\section{Optimization of the ansatz with symmetry breaking layers}\label{app:optimization_sym_brk}

When the ansatz contains symmetry-breaking layers, an optimization algorithm easily gets stuck in local minima. We here compare several different set-ups for the transverse field Ising model we have studied in Sec.~\ref{sec:ansatz}.
We show results from the transverse Ising model with different initial values and using the centered and uncentered Fisher matrix in Fig.~\ref{fig:tfi_brksym_compare}.
We can see that the result with $\{\phi_j\}$ initialized from $\mathcal{N}(0,\sigma^2)$ and using the centered Fisher matrix is the most natural. 
However, when $D=3,5$, the results with $\{\psi_j\}$ initialized from $\mathcal{N}(2\pi/D,\sigma^2)$ show better convergence. 

We have also found that using the uncentered Fisher matrix improves convergence for $D=4$ where the centered Fisher matrix failed to find an appropriate optimum. Our results suggest that the learning landscape of the VQEs with symmetry-breaking layers is rugged, especially when the parameters are not sufficient to describe the ground state accurately.

\bibliography{references}

\end{document}